\documentclass[reprint,amsmath,amssymb,prb,]{revtex4-2}
\usepackage{graphicx}
\usepackage{dcolumn}
\usepackage{bm}
\usepackage{hyperref}
\usepackage[utf8]{inputenc}
\usepackage[mathlines]{lineno}
\usepackage{xcolor}
\usepackage{algpseudocode}
\usepackage{multirow}
\begin{document}

\title{Thermodynamically Informed Priors for Uncertainty Propagation in First-Principles Statistical Mechanics}

\author{Derick E. Ober}
\author{Anton Van der Ven}
\date{\today}

\begin{abstract}
    This work demonstrates how first-principles thermodynamic research within a Bayesian framework can quantify and propagate uncertainties to downstream thermodynamic calculations. To address the issue of Bayesian prior selection, knowledge of zero kelvin ground states in the material system of interest is incorporated into the prior. The effectiveness of this framework is shown by creating a phase diagram for the FCC Zirconium Nitride system, including confidence intervals on phase boundary regions of interest.
\end{abstract}
\maketitle

\section{Introduction}
First-principles statistical mechanics \cite{de1994cluster,ceder1993derivation,van2018first} enables the prediction of finite temperature properties, such as alloy phase diagrams \cite{sanchez1991first,asta1992first,sluiter1996first,kohan1998computation,ozolicnvs2001large,van2002automating,van2002self,sluiter2006ab,zhou2006configurational,puchala2013thermodynamics,page2015phase,hao2016quaternary,goiri2016phase,hua2018first,
gunda2018first,gunda2018resolving,solomon2019stability,gunda2020understanding,cao2019computationally,natarajan2017first,natarajan2020crystallography,rahm2021tale}, magnetic order-disorder phenomena \cite{magnetic_order_disorder,decolvenaere2019modeling,kitchaev2020mapping,zuo2021magnetoentropic} and non-dilute diffusion coefficients. \cite{van2008nondilute,van2010vacancy,bhattacharya2011first,goiri2019role,kolli2021elucidating,kaufman2022cation}
This is generally achieved by first training a surrogate model \cite{sanchez1984generalized,de1994cluster,drautz2004spin,Mueller2009,thomas2013finite,thomas2018hamiltonians,thomas2019machine,drautz2019atomic,mueller2020machine,hart2021machine} on high-accuracy, zero kelvin data, obtained using Density Functional Theory (DFT) or one of its extensions \cite{lejaeghere2016reproducibility}, and then implementing the surrogate model within Monte Carlo simulations to calculate thermodynamic and kinetic properties at elevated temperatures. 
A variety of surrogate models have been developed to interpolate first-principles energies within Monte Carlo and molecular dynamics simulations. 
These include harmonic phonon Hamiltonians \cite{van2002effect,fultz2010vibrational}, alloy and magnetic cluster expansions \cite{sanchez1984generalized,de1994cluster,drautz2004spin,van2018first} and the atomic cluster expansion,\cite{drautz2019atomic} the latter serving as a rigorous foundation for modern machine learned interatomic potentials.\cite{deringer2021gaussian,musil2021physics} 

With rare exceptions \cite{KRISTENSEN2014, ALDEGUNDE2016}, first-principles statistical mechanics efforts have historically lacked transparent recognition and reporting of uncertainties. 
Each stage - generating data, training a model, calculating properties and post processing - involves approximations and numeric noise that introduce uncertainty into the final results. 
Quantifying this uncertainty is essential, and is increasingly feasible as computational resources become less expensive and more capable. 

For continuous thermodynamic quantities such as free energy and order-disorder transition temperatures, confidence intervals express uncertainty quantitatively. 
However, other thermodynamic properties are discrete or qualitative in nature; a system's set of ground states is one important example. 
A solid, such as an alloy, picks a small subset among an infinite number of microstates as having the lowest energy. 
When a surrogate model interpolates accurate zero kelvin data to predict finite temperature  phase stability, these ground states often remain stable at high temperatures. 
If the model predicts ground states that do not agree with the high accuracy data, it will produce a phase diagram that is un-supported by the data.
The phase diagram is then qualitatively incorrect, or qualitatively uncertain.

In general, there is only one true set of ground states. Assuming these ground states are known, a surrogate model should replicate them. 
In practice, changes to a surrogate model may or may not change the predicted ground states: this ground state problem describes a qualitative discontinuity in predicted ground states with respect to continuous changes in model parameters. This discontinuity can therefore be used to guide surrogate model selection, and reduce uncertainty in derived thermodynamic quantities. 

Bayesian frameworks are increasingly used to quantify and propagate uncertainties in multi-scale calculations of materials properties.\cite{Mueller2009,cockayne2010building,KRISTENSEN2014,ALDEGUNDE2016,honarmandi2019bayesian,honarmandi2019uncertainty} 
However, an ambiguous component of a Bayesian approach is the selection of Bayesian priors.\cite{Bishop2006} By restricting surrogate model priors to those replicating correct ground states, the ground state enforcement is formalized, and  the prior gains a physically-motivated interpretation. Once the prior distribution is defined, it is possible to quantify uncertainty in the surrogate model, and propagate the uncertainty to dependent calculations.\cite{Bishop2006} 

Here we develop Bayesian uncertainty quantification techniques that account for the ground state problem.
We then apply them to the alloy cluster expansion surrogate model.\cite{sanchez1984generalized,de1994cluster,van2018first} 
Alloy cluster expansions are rigorous and highly effective for studying thermodynamic properties derived from configurational disorder in multi-component crystals. 
At the same time, alloy cluster expansions are sufficiently simple to illustrate the principles described here. 
To provide context, the proposed methods are applied to quantify the uncertainty in the phase diagram of the ZrN$_x$ rocksalt phase, in which zirconium (Zr) atoms form an FCC sublattice while nitrogen atoms (N) occupy octahedrally coordinated interstitial sites. 
This compound can tolerate high concentrations of nitrogen vacancies, thereby leading to configurational disorder on the nitrogen sublattice. 
At low temperatures, the nitrogen and vacancies order at stoichiometric compositions. 
These ordered phases constitute the ground states of the ZrN$_x$ system and play a defining role in determining the topology of the temperature versus nitrogen composition phase diagram.

\section{Method}
\subsection{Surrogate model and ground states}
\subsubsection{Convex hull and ground state orderings}
\begin{figure}
    \centering
\includegraphics[width=1\linewidth,height=1\textheight,keepaspectratio]{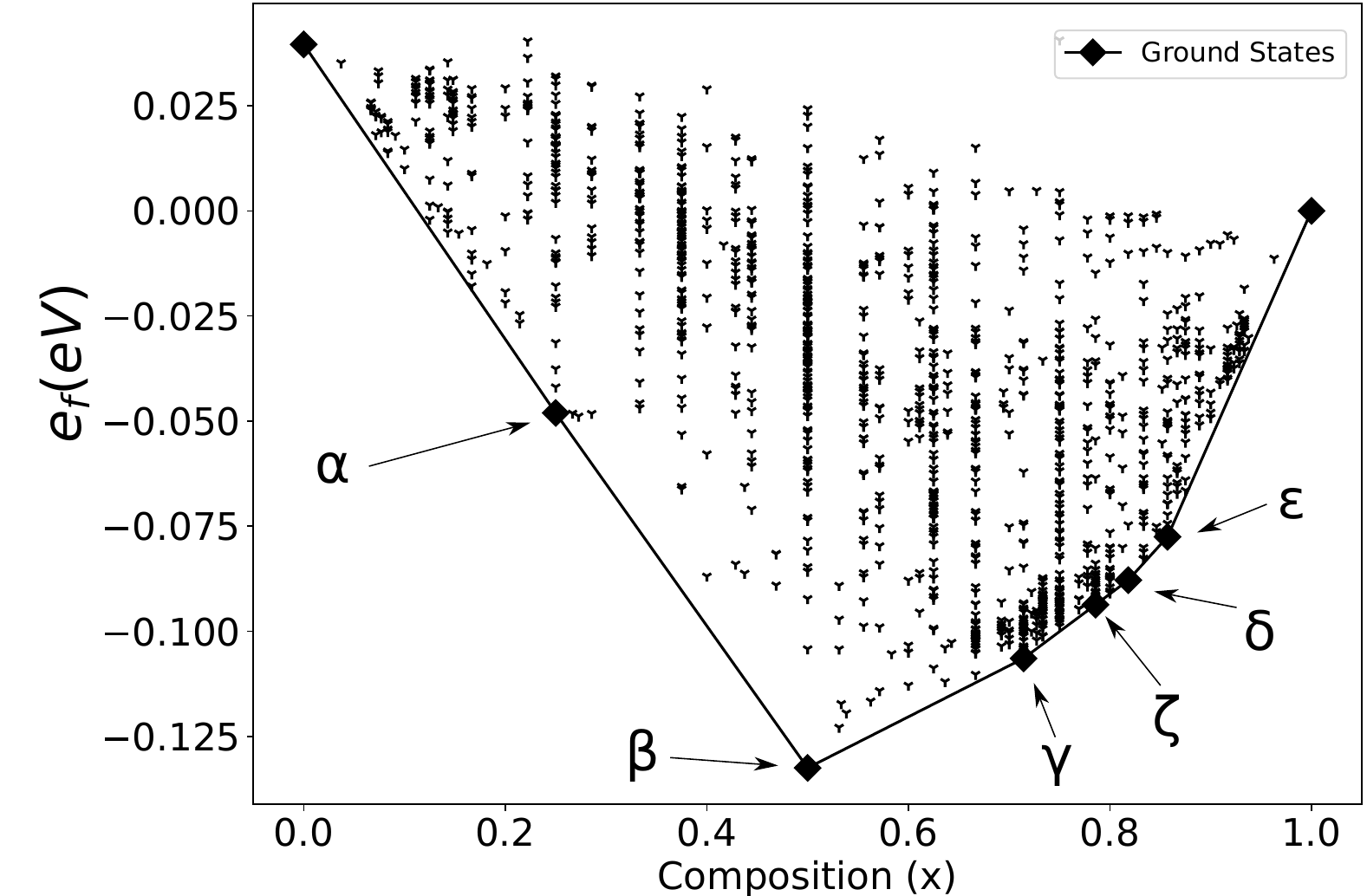}
    \caption{DFT formation energies and lower convex hull for FCC ZrN$_x$. All points correspond to a unique configuration of N atoms on interstitial octahedral sites. Diamonds are ground state configurations, labeled by Greek letters, and the line connecting them is the lower convex hull.}
    \label{fig:convex_hull}
\end{figure}

An alloy's ground states are the chemical orderings over sites in a parent crystal structure \cite{kolli2020discovering} that minimize the energy of the alloy at zero kelvin. 
When minimizing the energy of an alloy, it is most convenient to work with formation energies.
For ZrN$_x$ with the rocksalt structure, these are defined as 
\begin{equation}\label{formation_energy_equation}
    e_{f}(\vec{\sigma}) = e[\vec{\sigma}] - ((1-x)e[\text{Zr}] + (x)e[\text{ZrN}])
\end{equation}
where $e[\Vec{\sigma}]$ is the energy per primitive unit cell of some arrangement of N atoms on octahedral interstitial sites of FCC Zr, labeled as $\Vec{\sigma}$, the $x=0$ reference state is HCP Zr without N atoms, and the $x=1$ reference state is the rocksalt ZrN structure, where all octahedral interstitial sites are occupied by N atoms. 
Figure \ref{fig:convex_hull} shows formation energies of 1297 N-vacancy configurations over the FCC Zr interstitial sites, as calculated with density functional theory (See Results section for details). 
Diamonds on the lower convex hull are the formation energies of the ground state configurations. 

\subsubsection{Cluster Expansion}\label{cluster_expansion_section}
The alloy cluster expansion is a surrogate model describing the energy of a multicomponent crystal as a function of its degree of order.\cite{sanchez1984generalized,de1994cluster} 
In the case of a binary solid, occupation variables $\sigma_j$ are assigned to each crystal site $j$, taking a value of -1 or +1 depending on the occupation of that site. A specific atomic configuration over the $M$ unit cells of the crystal is specified by the collection of all occupation variables $\vec{\sigma} = \{\sigma_1, \sigma_2, ..., \sigma_{M\times s}\}$, where $s$ is the number of sites per unit cell.
The energy of this arrangement can then be expressed as \cite{sanchez1984generalized}
\begin{equation}\label{cluster_hamiltonian}
    E_f(\vec{\sigma}) = MV_0+ \sum_{c} V_{c}\phi_{c}(\vec{\sigma})
\end{equation}
where $\phi_{c}$ are cluster basis functions defined as the products of occupation variables belonging to clusters: 
\begin{equation}\label{cluster_function}
    \phi_c(\vec{\sigma}) = \prod_{j\in c}\sigma_j
\end{equation}
A cluster is a collection of sites, which may be a point (one site), pair (two sites), triplet (three sites) etc., and is indexed by $c$. 
Meanwhile, $V_c$ are chemistry-dependent expansion coefficients that quantify the extent with which each basis function $\phi_c$ contributes to the total energy of the crystal.
Crystal symmetry places restrictions on the expansion coefficients; if any two basis functions $\phi_c$ and $\phi_{c'}$ can be mapped onto each other by one of the crystal space group operations, their corresponding coefficients, $V_c$ and $V_{c'}$, must be equal. 
Cluster functions that are symmetrically equivalent can be grouped into orbits $\Omega_c$ of basis functions. All cluster functions within the same orbit have a common expansion coefficient $V_c$. 
In terms of orbits, the crystal energy can be expressed as \cite{sanchez1984generalized}
\begin{equation}
    E_f(\vec{\sigma}) = MV_0+ \sum_{\Omega_c \in \Lambda} V_{c}\sum_{c' \in \Omega_{\alpha}}\phi_{c'}(\vec{\sigma})
\end{equation}
where $\Lambda$ is the set of all orbits $\Omega_{c}$. The energy can next be normalized by the number of primitive unit cells $M$ in the crystal 
\begin{equation}\label{final_clex_hamiltonian}
    e_f(\vec{\sigma})=\frac{E_f(\vec{\sigma})}{M} = V_0+ \sum_{\Omega_c \in \Lambda} V_{c}m_{c}\langle\phi_{c}(\vec{\sigma})\rangle
\end{equation}
where the sum over basis functions belonging to a common orbit is replaced by a correlation function defined as \cite{sanchez1978fee,sanchez1982ising,sanchez1984generalized}
\begin{equation}\label{eq:correlation_definition}
    \xi_{c}(\vec{\sigma})=\langle \phi_{c}(\vec{\sigma})\rangle = \frac{\sum_{{c'} \in \Omega_{c}} \phi_{c'}(\vec{\sigma})}{Mm_{c}}
\end{equation}
The $m_c$ are the number of symmetrically equivalent cluster functions per primitive unit cell in orbit $\Omega_c$.

The expansion coefficients $V_\alpha$ are referred to as Effective Cluster Interactions (ECIs) \cite{sanchez1984generalized,de1994cluster} and must be determined by fitting to first-principles training data. 
The correlation functions for a particular ordering $\Vec{\sigma}$ can be collected as a vector $\Vec{\xi}^{\mathsf{T}}(\Vec{\sigma}) = (1,\xi_1(\Vec{\sigma}),\xi_2(\Vec{\sigma}),\xi_3(\Vec{\sigma}),...)$, where the first entry, $\xi_0 = 1$, corresponds to the correlation function of the empty cluster having coefficient $V_0$. 
By similarly collecting the ECI of each orbit in a vector $\Vec{w}^{\mathsf{T}} = (V_0, V_1 m_1, V_2 m_2, V_3 m_3, ...)$, the energy of a configuration $\Vec{\sigma}$ can be expressed as the scalar product 
\begin{equation}\label{eq:correlation_multiply}
    e_f (\Vec{\sigma}) = \Vec{\xi}^{\mathsf{T}}(\Vec{\sigma})\vec{w}=\sum_{c}^{K} \xi_c(\Vec{\sigma}) V_c m_c 
\end{equation}
Using notation common to statistical learning,\cite{Bishop2006} the feature vector $\Vec{\xi}^{\mathsf{T}}(\Vec{\sigma})$ for a configuration $\Vec{\sigma}$ can form a row of a feature matrix $X$. 
The energies of $N$ configurations can then be represented as a matrix multiplication 
\begin{equation}\label{linear_model_cluster_expansion}
    \vec{t} = X\vec{w}
\end{equation}
where $\vec{t}$ collects the formation energies of the $N$ configurations.  
In a finite-sized binary crystal with $M\times s$ sites, there are $2^{M\times s}$ possible configurations $\vec{\sigma}$; there are also $2^{M\times s}$ possible clusters. 
When $M\to\infty$, the number of configurations and the number of clusters and corresponding basis functions also becomes infinite. 
If all basis functions are used, the cluster expansion model is capable of exactly predicting the energy of each configuration.\cite{sanchez1984generalized,de1994cluster} 
In practice, the basis set must be truncated.
Since chemical interactions in solids are usually local, a small number of basis functions are often sufficient to replicate the majority of the observed data.

A cluster expansion surrogate model, defined by $\Vec{w}$, can predict the energy of any atomic configuration, including configurations $\Vec{\sigma}$ in supercells that are too large for high accuracy electronic structure methods. 
Monte Carlo simulations leverage this to calculate the energies of millions of configurations sampled in large supercells. 
The dramatic speedup enables the exploration of possible new  ground states using simulated annealing as well as the calculation of thermodynamic quantities such as the composition dependence of chemical potentials, free energies, and ultimately the full phase diagram of the alloy.

\subsubsection{The Ground State Problem}\label{ground_state_problem_section}
A truncated cluster expansion should replicate first-principles formation energies with high fidelity and simultaneously predict the correct ground states.
Cluster expansions that fail to predict correct ground states yield finite temperature phase diagrams that qualitatively disagree with the higher accuracy first-principles electronic structure method. 
The ability of a cluster expansion to predict a particular set of ground states is sensitive to the level of truncation.
Furthermore, for each level of truncation, there are well-defined domains within the ECI vector space, $\vec{w}$, in which the cluster expansion predicts a fixed set of ground states.\cite{geometric_interpretation,ducastelle1993order,inden2001atomic}
We illustrate this for atom-vacancy disorder on a triangular lattice. 
Note that all Greek phase labels in this section are not related to the phase labels used to refer to the ground states of ZrN$_x$ in other sections.

\begin{figure}[!h]
    \centering
    \includegraphics[width=1\linewidth,height=1\textheight,keepaspectratio]{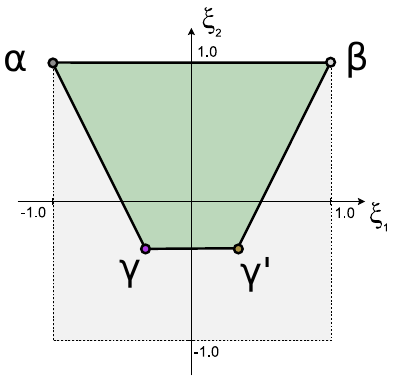}
    \caption{Triangular lattice correlation space for point and first nearest neighbor interaction. Vertices of a polytope in correlation space describe the possible ground states for the chosen correlation basis.}
    \label{fig:triangular_lattice_correlations}
\end{figure}

\begin{figure}[!h]
    \centering
\includegraphics[width=1\linewidth,height=1\textheight,keepaspectratio]{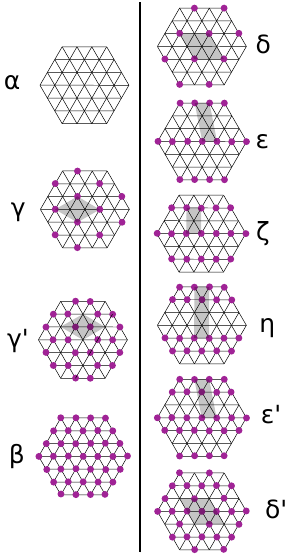}
    \caption{Possible ground state orderings when accounting for first neighbor interactions (left), and both first and second nearest neighbor interactions (all). Structures are ordered from lowest composition at the top to highest composition at the bottom.}
    \label{fig:orderings}
\end{figure}

For every possible ordering $\Vec{\sigma}$ on a triangular lattice (with $\sigma_{j}=+1$/$-1$ if site $j$ is occupied/vacant), there is a corresponding vector of correlations $\Vec{\xi}(\vec{\sigma}) = (1,\xi_1(\vec{\sigma}), \xi_2(\vec{\sigma}),\dots)$ whose values can be calculated using Eq. \ref{eq:correlation_definition}.
For exposition purposes, we initially limit ourselves to the correlations of the point and nearest neighbor pair cluster on the triangular lattice.
Let $\xi_1$ be the correlation function for the point cluster and  $\xi_2$ the correlation function for the nearest neighbor pair cluster.
Since correlation functions are averages of products between +1's and -1's, their values are restricted to the interval $[-1,1]$.
The parent crystal structure, however, may impose additional constraints on the values of the correlations. 
For example, the geometry of the triangular lattice causes ``frustration",\cite{ducastelle1993order} which restricts correlations to a narrower interval. 
Figure \ref{fig:triangular_lattice_correlations} shows the convex polytope that encompasses the correlations of all physically realizable orderings on the triangular lattice in the two-dimensional correlation space of $\xi_1$ and $\xi_2$.\cite{ducastelle1993order}
The vertices of the polytope in Figure \ref{fig:triangular_lattice_correlations} correspond to the $x=0$ and $x=1$ pure states and the $\sqrt{3}a\times\sqrt{3}a$ superlattice orderings at $x=1/3$ and $x=2/3$. These four orderings are depicted on the left hand side of Figure \ref{fig:orderings}. 

It can be shown that among all possible configurations, those with correlations at polytope vertices in correlation space extremize the energy.\cite{ducastelle1993order,inden2001atomic} 
Therefore, they are the only possible ground states of the truncated cluster expansion. 
In this particular example, only the four vertex orderings of Figure \ref{fig:triangular_lattice_correlations} are possible ground states for a truncated cluster expansion consisting of the point and nearest neighbor pair terms. 
If a system has a known ground state ordering that is not a vertex in the chosen correlation space, additional basis functions must be added to the truncated cluster expansion until the desired ground state ordering appears as a polytope vertex in correlation space.\cite{ducastelle1993order,inden2001atomic} 

\begin{figure}[!h]
    \centering    \includegraphics[width=1\linewidth,height=1\textheight,keepaspectratio]{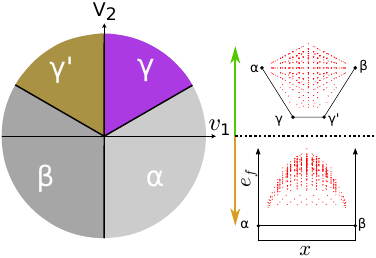}
    \caption{Triangular lattice point and first nearest neighbor ECI space: different regions in ECI space stabilize different ground states. Ground states are only determined by direction, not by vector magnitude. Projecting along $v_1$ by varying $\mu$ reveals two distinct convex hulls: one for $V_2>0$, another for $V_2 < 0$.} 
    \label{fig:triangular_lattice_eci}
\end{figure}

\begin{figure}
    \centering
\includegraphics[width=1\linewidth,height=1\textheight,keepaspectratio]{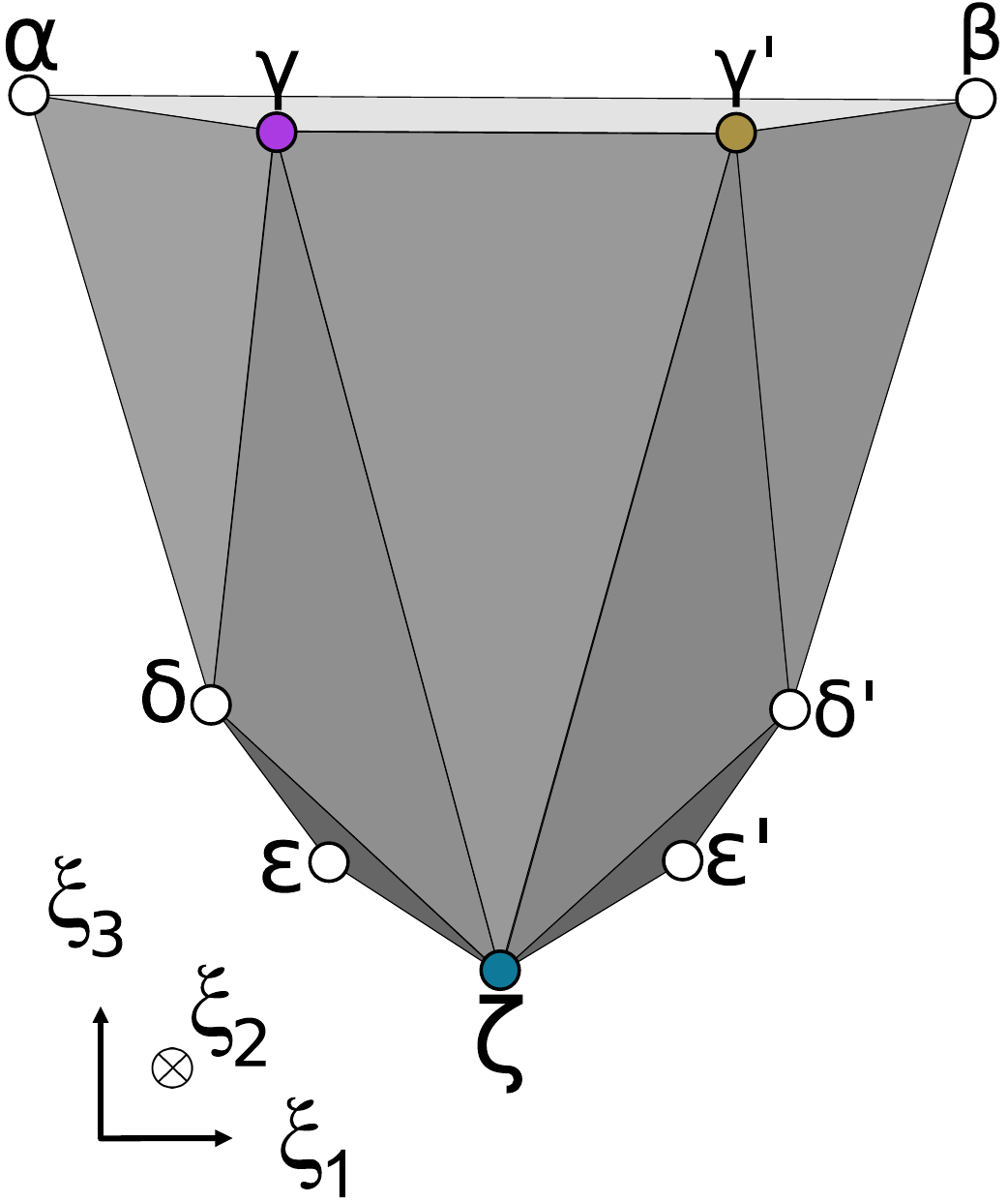}
    \caption{Vector space of point, first and second nearest neighbor correlations ($\xi_1,\xi_2, \xi_3$) and corresponding polytope for the triangular lattice. Vertices of the polytope describe possible ground states. Adding more interaction dimensions increases the number of polytope vertices, and therefore the number of possible ground states. The vertex for the $\eta$ ordering is hidden behind the polytope in this image. The $\gamma$, $\gamma$' and $\zeta$ vertices are colored to match their corresponding stability regions in Figures \ref{fig:triangular_lattice_eci}, \ref{fig:3d_eci_rays_with_projection}. }
    \label{fig:3d_corr_polytope}
\end{figure}

Since the cluster expansion as formulated in Eq. \ref{eq:correlation_multiply} is a scalar product between $\vec{\xi}$ and $\vec{w}$ (with $\Vec{w}^{\mathsf{T}} = (V_0, V_1 m_1, V_2 m_2, V_3 m_3, ...)$), the ECI reside in the dual space of the correlation vector space.
The polytope in correlation space, $\vec{\xi}$, can therefore be converted to a phase diagram map in ECI space, $\vec{V}$.\cite{ducastelle1993order,inden2001atomic} 
An example is shown in Figure \ref{fig:triangular_lattice_eci} for the truncated cluster expansion of the triangular lattice containing only the point and nearest neighbor pair terms. 
Each domain in this phase diagram represents ECI values that stabilize a particular ground state. It should be noted that the ground state for a given ECI vector is only determined by ECI vector direction, not magnitude. Scaling the magnitude of the ECI sets the energy scale of the model, which enlarges or shrinks the distance between the highest and lowest energy structures. Qualitatively, this causes the formation energy points to ``breathe" without changing the ground states.

The addition of a second nearest-neighbor interaction to a cluster expansion for the triangular lattice expands the correlation space to three dimensions. The corresponding three dimensional polytope is shown in Figure \ref{fig:3d_corr_polytope}. 
Notice that the trapezoid polytope of Figure \ref{fig:triangular_lattice_correlations} appears as the top face of the new polytope in three dimensions. 
Expanding the space of possible interactions adds additional vertices, corresponding to new possible ground state orderings of the truncated cluster expansion. 
The ground states of a triangular lattice cluster expansion that contains both nearest neighbor and second nearest neighbor interactions now includes all orderings depicted in Figure \ref{fig:orderings}.

\begin{figure*}
    \centering
    \includegraphics[width=1\linewidth,height=1\textheight,keepaspectratio]{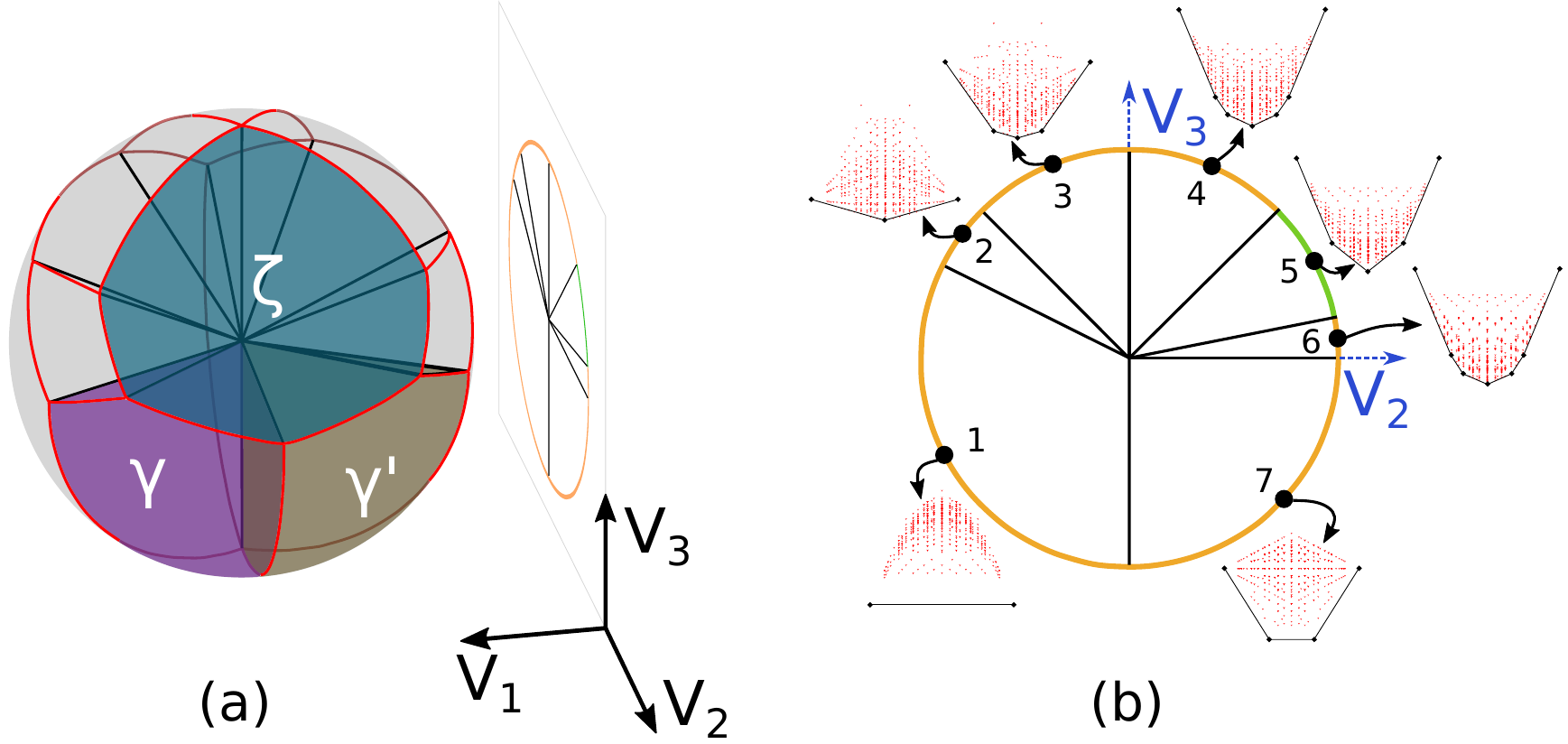}
    \caption{(a): ECI dual space of Figure \ref{fig:3d_corr_polytope}, including point, first nearest neighbor and second nearest neighbor ECIs ($V_1$,$V_2$,$V_3$ respectively). Each arc volume in the unit sphere (a) corresponds to a single stable ordering. For reference, regions in which the orderings $\gamma$, $\gamma$' and $\zeta$ are stable are colored in purple, brown and blue respectively. Projecting the unit sphere and boundaries in (a) along $V_1$ gives ray-bounded ``cone"-like regions in $V_2$, $V_3$ space (b), where each cone between rays corresponds to a region of shared convex hull vertices (shared ground states). An ECI vector was taken from the center of each cone to create a prototypical plot of formation energies and a convex hull for that cone. For example purposes, cone 5 will be used as the ``correct" cone, and is colored green to represent this. All other cone arcs are colored orange, to represent the fact that they are ``incorrect", and that there can be at most one cone that replicates ground states perfectly. }
    \label{fig:3d_eci_rays_with_projection}
\end{figure*}

The inclusion of a second nearest neighbor correlation, $\xi_3$, and its corresponding ECI, $V_3$, extends the ECI ground state map to three dimensions. 
This is depicted in Figure \ref{fig:3d_eci_rays_with_projection}, where distinct ground state domains in ECI vector space are shown as multi-faceted cones that extend outward from the origin.
Their intersection with the unit sphere in ECI space are shown with the red lines.
Figure \ref{fig:3d_eci_rays_with_projection} is analogous to Figure \ref{fig:triangular_lattice_eci}, where each cone corresponds to a particular ordering. 
For example, the $x=\frac{1}{3}$ and $x=\frac{2}{3}$ orderings of $\gamma$ and $\gamma '$ again correspond to the purple and brown regions respectively, while the new $x=\frac{1}{2}$ $\zeta$ phase corresponds to the blue region of Figure \ref{fig:3d_eci_rays_with_projection}, and the blue vertex dot in Figure \ref{fig:3d_corr_polytope}. 
Again, when a given ECI vector lies in one of the cones, the structure corresponding to that region will have the lowest possible formation energy for that choice of ECI vector.

Ground state optimization is most conveniently done in the grand canonical ensemble where the chemical potential, $\mu$, as opposed to the composition, $x$, is held constant.
This is the ensemble of an open system in equilibrium with a reservoir of atoms at a constant chemical potential $\mu$. 
At constant $\mu$, the ground states of the system minimize the grand canonical energy defined as 
\begin{equation}
\psi(\Vec{\sigma}) = e_f (\Vec{\sigma}) - \mu x,
\label{eq:grand_canonical_energy}
\end{equation}
This quantity can also be expressed as a scalar product between a correlation vector, $\vec{\xi}$, and a vector of ECI, $\vec{v}$, upon inserting Eq. \ref{eq:correlation_multiply} into Eq. \ref{eq:grand_canonical_energy} and using $x=1/2(1+\xi_1)$ to yield
\begin{equation}
    \psi(\Vec{\sigma}) = \Vec{\xi}^{\mathsf{T}}(\Vec{\sigma})\vec{v}
\end{equation} 
In this expression, $v_0=V_0-\mu/2$ and $v_1=V_1-\mu/2$ while $v_{c}=m_{c}V_{c}=w_{c}$ for $c > 1$. 
An advantage of working within the grand canonical ensemble is that the full set of ground states for a specific vector of ECI can be surveyed by sweeping over different chemical potential values $\mu$.
In the ground state map of Figure \ref{fig:triangular_lattice_eci}, for example, the horizontal axis corresponding to the point cluster interaction, $v_1=V_1-\mu/2$, can be traversed from negative to positive values by varying $\mu$. 
For a fixed value of $V_2$, a sweep over $v_1$ tells us the possible ground states of the truncated cluster expansion over the whole composition range. 
When $V_2 < 0$, the only ground states are the $x=0$ and $x=1$ configurations. When $V_2 > 0$, all four orderings at $x=0$, $x=1/3$, $x=2/3$ and $x=1$ are ground states of the model. The cases of $V_{2}<0$ and $V_2 >0$ are the two qualitatively distinct convex hulls when considering only the first nearest neighbor interaction.

The values of the ECI corresponding to a fixed {\it set} of ground states also form a cone in ECI space. 
This can be illustrated for the triangular lattice cluster expansion containing both the first and second nearest neighbor interactions, $V_2$ and $V_3$.
Projecting the ECI ground state map (Figure \ref{fig:3d_eci_rays_with_projection}(a))  along the $V_1$ axis, which is equivalent to sweeping over the chemical potential, generates cones in $V_2$, $V_3$ space (Figure \ref{fig:3d_eci_rays_with_projection}(b)) corresponding to different ground state sets.
As shown in Figure \ref{fig:3d_eci_rays_with_projection}(b), a triangluar lattice cluster expansion with first and second nearest neighbor interactions, $V_2$ and $V_3$, has seven qualitatively distinct convex hulls. 
All ECI vectors that fall within a given cone will form a convex hull with the same ground states. 
Moving the ECI vector within a cone will change predicted energies, but the predicted set of ground states will only change when an ECI vector crosses from one cone to another. 
Representative convex hulls for each cone are shown in Figure \ref{fig:3d_eci_rays_with_projection}(b) and were generated by taking an ECI vector from the center of each numbered cone (on the unit circle), and evaluating the energies of a large number of symmetrically distinct orderings on the triangular lattice. 
The member ground states for each qualitative convex hull are summarized in Table \ref{table:ground_state_sets}. 

\begin{table}[]
    \centering

    \begin{tabular}{|c|p{3cm}|}
 \hline
 Cone Index & Ground States\\
 \hline
 1 & $\alpha,\beta$ \\
 2 &$\alpha,\eta,\beta$  \\
 3 & $\alpha,\varepsilon,\eta,\varepsilon ', \beta$ \\
 4 & $\alpha,\delta, \varepsilon,\zeta, \varepsilon',\delta',\beta$ \\
 5 & $\alpha,\delta,\zeta,\delta',\beta$ \\
 6 & $\alpha,\delta,\gamma,\zeta,\gamma',\delta',\beta$ \\
 7 & $\alpha,\gamma,\gamma',\beta$ \\
 \hline
\end{tabular}

    \caption{Unique ground state sets for the first and second nearest neighbor triangular lattice cluster expansion.}
    \label{table:ground_state_sets}
\end{table}

\subsection{Bayesian Surrogate model parameterization} 
The expansion coefficients of a truncated cluster expansion, $\vec{w}$, can be trained to the formation energies of a large number of orderings as calculated with a first-principles electronic structure method such as density functional theory. 
A Bayesian approach to model parameter determination is desired as it enables a formalized incorporation of prior knowledge and biases. 
This section first reviews Bayesian approaches to constructing surrogate models using training data and then introduces a new Bayesian approach to generate cluster expansions and enable uncertainty quantification by incorporating prior knowledge about ground states.

\subsubsection{Bayesian Uncertainty Quantification, Propagation}

Bayesian statistics interprets probability as a degree of belief, and is underpinned by Bayes' Theorem:
\begin{equation}\label{bayes_theorem}
P(A|B) = \frac{P(B|A)P(A)}{P(B)}    
\end{equation}
Bayes' theorem arises from the product rule of probability, where ``prior" knowledge of a random variable $A$ can be updated by observed data $B$ to produce a ``posterior" probability distribution on $A$ that accounts for prior knowledge and observed data. 
As applied to a truncated cluster expansion Hamiltonian, Bayes' Theorem becomes:

\begin{equation}\label{bayes_theorem_contextualized}
P(\vec{w}|\vec{t},X) = \frac{P(\vec{t}|X,\vec{w},\beta)P(\vec{w})}{P(\vec{t}|X,\beta)}
\end{equation}
In this relation, $P(\vec{w}|\vec{t},X)$ represents a posterior probability distribution for the ECI, $\vec{w}$, of a cluster expansion given first-principles formation energies for $N$ different orderings, as collected in the vector $\vec{t}$.
The feature matrix $X$ collects the correlation functions for the basis function of the cluster expansion, Eq. \ref{eq:correlation_definition}, evaluated for each of the $N$ configurations.
Each row in $X$ corresponds to a particular ordering on the lattice and each column of $X$ corresponds to a particular basis function of the cluster expansion.
The term $P(\Vec{w})$ is the ``prior" distribution on linear model coefficients $\Vec{w}$, and should encode expert knowledge of cluster expansion behavior. 
Values of $\Vec{w}$ that are in agreement with expected behavior are assigned a greater probability than values that are not in agreement.
The likelihood function $P(\vec{t}|X,\vec{w},\beta)$ updates the prior with observations $\Vec{t}$, the first-principles training data. 
Conceptually, the likelihood can be interpreted as answering the question: ``If it is assumed that $\vec{w}$ is the correct model and this model predicts energies through $X\vec{w}$, what is the probability of observing the data $\vec{t}$ considering that the numerical accuracy of the first-principles method that generated the data has a precision $\beta$ (defined as the inverse of the variance, $\sigma_L^2$, on the training data)?" 
(The use of the letter $\beta$ as the likelihood precision has no relation to the use of $\beta$ as a label for material phases.) 
The posterior distribution $P(\vec{w}|\vec{t},X)$ is the product of the prior and likelihood, and assigns a higher probability to models $\Vec{w}$ that are in agreement with both prior knowledge and observed data. 
The denominator in equation \ref{bayes_theorem_contextualized} normalizes the posterior distribution so that it is a true probability distribution. 

If additional calculations depend on the surrogate model, the uncertainty can be ``propagated" through those calculations. 
This produces another posterior distribution, encoding uncertainty for the calculated quantities. 
Propagation to a calculated quantity $Q$ is performed by marginalizing across model parameters $\vec{w}$:
\begin{equation}\label{propagation_integral}
P(Q) = \int P(Q|\vec{w})P(\vec{w}|\vec{t},X)d\vec{w}    
\end{equation}
where $Q$ is some quantity (e.g. predicted free energy, order-disorder transition temperature, etc.). 
The term $P(Q|\Vec{w})$ is the probability of observing each possible $Q$ given a specific selection of $\Vec{w}$. 
For example, in first-principles statistical mechanics, $P(Q|\Vec{w})$ can be understood as the uncertainty in thermodynamic calculations for a given vector of ECI $\Vec{w}$. 
Generally, it is the probability distribution across all values of $Q$ that are obtainable within the space of ECI $\Vec{w}$ and this term is weighted by the posterior probability of observing each $\Vec{w}$. 
The product of these two distributions forms the integrand, and is calculated across all values of $\Vec{w}$ to produce $P(Q)$, the probability distribution across all $Q$.

\subsubsection{Bayesian Model Fitting: the traditional approach}
\label{sec:bayesian_fitting}
The cluster expansion, before truncation, can describe the energy of an alloy exactly as it is an expansion in terms of a complete and orthonormal basis.\cite{sanchez1984generalized}
However, due to numerical convergence limitations of the first-principles DFT method that is used to generate the training data $\vec{t}$, equation \ref{linear_model_cluster_expansion} should be rewritten as
\begin{equation}\label{un_truncated_Bayesian_cluster_expansion}
    \vec{t} = X\vec{V} + \vec{\epsilon}.
\end{equation}
where $\vec{\epsilon}$ derives solely from DFT numerical uncertainty in the calculation of DFT energies (for a particular approximation of DFT such as LDA or GGA).

A common assumption is that the elements of $\vec{\epsilon}$ are drawn from a zero-centered, independently and identically distributed (iid) Gaussian, $\epsilon_i \thicksim \mathcal{N}(0,\beta^{-1})$, where the precision, $\beta$ is the inverse of the expected variance $\sigma_L^2$ on each training datum $t_i$ due to numerical convergence limitations of the first-principles DFT method. 
The value of $\beta$, therefore, depends on the convergence thresholds used in performing the DFT calculations.
The noise defines the functional form of the likelihood function $P(\vec{t}|X,\vec{V},\beta)$ of Bayes' Theorem, Equation \ref{bayes_theorem_contextualized}. 
In the assumption of iid Gaussian noise, the likelihood function is a product of univariate Gaussians for each observed data point:
\begin{equation}
P(\vec{t}|X, \vec{w}, \beta) = \prod^{N}_{i} \mathcal{N} (t_i|X_i\vec{w},\beta^{-1})
\label{eq:likelihood_function}
\end{equation}
If the iid assumption does not hold, the likelihood distribution would be defined as a multivariate Gaussian with nonzero off-diagonal elements in its covariance matrix.\cite{Bishop2006}

A common choice for the prior distribution $P(\vec{w})$ of Equation \ref{bayes_theorem} is a normal distribution with a mean $\vec{\mu}$ and precision $\alpha$ (the inverse variance on the ECI, not to be confused with the name of an ordered phase). 
The uncertainty on ECI is then assumed to be iid, reducing the full prior $P(\vec{w})$ to a product of individual priors for each ECI. 
This gives a product of normal distributions, where each ECI is defined by an inverse variance $\alpha_k$ and a mean $\mu_k$ for the $k^{th}$ ECI.
\begin{equation}
P(\vec{w}) = \prod^{K}_{k} \mathcal{N}(w_k|\mu_k,\alpha_k^{-1})
\label{eq:prior_distribution}
\end{equation}
If the ECI are correlated, they cannot be expressed as a product of univariate distributions and must instead be described by a full covariance matrix (and possibly higher order moment tensors). 

Many regression methods choose a mean zero prior ($\mu_k=0$) by default.\cite{Mueller2009,cockayne2010building,nelson2013cluster,nelson2013compressive,KRISTENSEN2014,ALDEGUNDE2016} 
One common approach is cross validated ridge regression, which is a form of regularized regression. \cite{hoerl1970ridge} 
In a Bayesian context, ridge regression can be interpreted as having a mean zero prior, $\mu_k=0$, with some standard deviation, $1/\sqrt{\alpha}$, related to the regularization coefficient, $\lambda = \alpha/\beta$, selected by cross validation. 
It is also common to choose other forms for the prior, including a Laplacian distribution, a uniform distribution, or a more exotic distribution such as those containing hierarchical hyper priors.\cite{Bishop2006,Mueller2009,nelson2013cluster,nelson2013compressive,ALDEGUNDE2016}
Regardless of the selected distribution, the set of cluster basis functions will always be a finite truncation of the full cluster basis. 
The exclusion of some cluster basis functions is equivalent to stating that the prior distribution for the corresponding ECI are Dirac delta distributions centered at zero.\cite{Mueller2009,cockayne2010building}
When the prior precision, $\alpha$, on the non-truncated ECI goes to zero, the prior distribution becomes uniform and the posterior maximum {\it a posteriori} (MAP) estimate of $\vec{w}$ (corresponding to the ECI vector with the maximum posterior probability) reduces to the least squares solution.

\subsubsection{Imposing prior knowledge about the ground states}
\label{sec:Bayesian_ground_state_priors}
\begin{figure*}
    \centering
\includegraphics[width=1\linewidth,height=1\textheight,keepaspectratio]{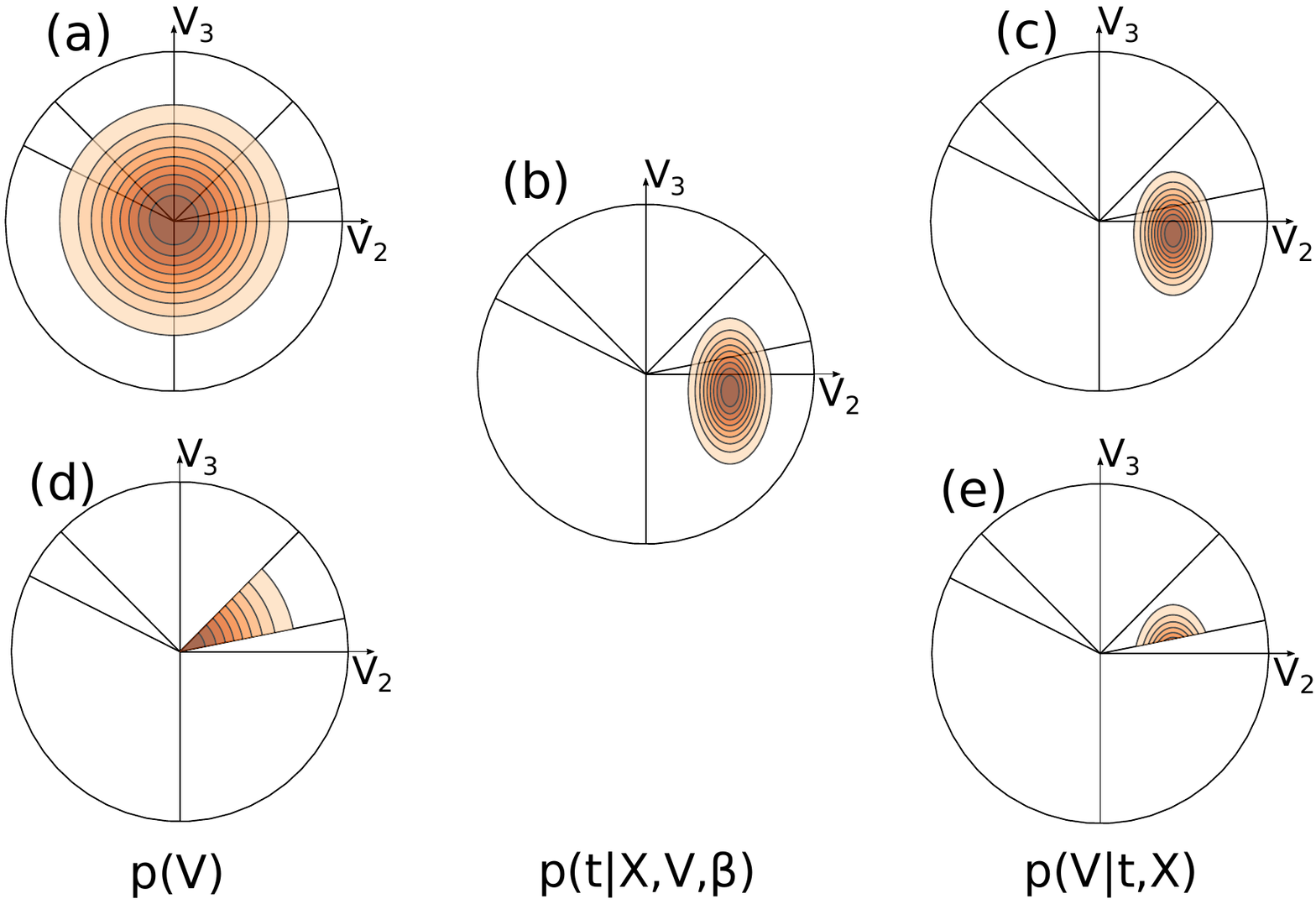}
    \caption{A mean-zero, Gaussian prior places high probability on models with small coefficients, and places equal probability density across all possible convex hull cones (a). The likelihood function places high probability on models that support observed data (b). The product of the mean-zero, Gaussian prior and the likelihood function produces a posterior distribution (c).  A prior can instead be restricted to a single ground state cone, where probability is highest within the ``true" cone (d). The product between the restricted cone prior and the likelihood function produces another possible posterior distribution (e). }
    \label{fig:prior_posterior_cartoon}
\end{figure*}

The main contribution of this work is a Bayesian framework to determine cluster expansions guided by knowledge about the ground states of the alloy. 
As described in the previous section, past Bayesian approaches \cite{mueller2020machine,cockayne2010building,nelson2013cluster,nelson2013compressive} assume a mean-zero prior distribution. 
When all the ECI are zero, every possible ordering in the crystal is degenerate in energy and no prior knowledge about the ground states of the system is exploited.
To illustrate this more concretely, Figure \ref{fig:prior_posterior_cartoon}(a) shows a contour plot of a Gaussian mean-zero prior distribution in the ECI space of a triangular lattice cluster expansion, spanned by the nearest and next nearest neighbor interactions, $V_2$ and $V_3$. 
As is evident in Figure \ref{fig:prior_posterior_cartoon}(a), a Gaussian mean-zero prior distribution has a finite probability within every ground state cone. 
Figure \ref{fig:prior_posterior_cartoon}(b) schematically illustrates a likelihood function in the same ECI space, which can be constructed once training data $\vec{t}$ has been collected. 
The product of the likelihood and the prior functions generates the posterior distribution, shown schematically as a contour plot in Figure \ref{fig:prior_posterior_cartoon}(c). 
Similar to the prior distribution and likelihood, this posterior distribution will in general straddle multiple ground state cones and ECI vectors sampled from such a distribution will not always predict the same ground states. 
Furthermore, the maximum {\it a posteriori} (MAP) ECI vector (corresponding to the ECI vector with the maximum posterior probability), may not reside within the ground state cone corresponding to that of the physical system of interest. 
In such a scenario, the temperature versus composition phase diagram of the alloy calculated with the MAP ECI vector will be topologically incorrect due to erroneous ground state predictions. 

An alternative to the Gaussian mean-zero prior distribution is desired in situations where the correct ground states are known {\it a priori} with a high degree of confidence.
The simplest prior in this scenario is a uniform distribution within the ``true" ECI cone, and zero outside of this cone. However, the cones are unbounded in the radial direction, which cannot be normalized and is therefore not a probability distribution. In order to make it a bounded distribution, the radial probability can be regularized, just as in the Gaussian mean-zero case. 
Figure \ref{fig:prior_posterior_cartoon}(d) illustrates this schematically within the ECI space of the truncated triangular lattice cluster expansion. 
A prior distribution that is bounded within the ``true" ground state cone and zero elsewhere guarantees that the posterior distribution, obtained by multiplying the prior with the likelihood function (Figure \ref{fig:prior_posterior_cartoon}(b)) only has a finite probability within the ``true" ground state cone. 
This is illustrated in Figure \ref{fig:prior_posterior_cartoon}(e), which shows that the posterior distribution is zero outside the ``true" ground state cone, but retains the qualitative shape of the likelihood function within the ground state cone. 
Any ECI vector sampled from this posterior distribution will be guaranteed to predict the correct ground states. 
Furthermore, since every sampled ECI vector from such a distribution has the same ground states, it becomes meaningful to calculate uncertainty bounds on order-disorder transition temperatures. 

The distributions schematically plotted in Figure \ref{fig:prior_posterior_cartoon} represent the two extremes of prior ground state knowledge: 
in Figure \ref{fig:prior_posterior_cartoon}(a) and (c), no knowledge about the ground states is used, while in Figure \ref{fig:prior_posterior_cartoon}(d) and (e) the resulting posterior distribution reflects a 100$\%$ certainty about the ground states.
In practice, there may be occasions where these two extremes are too restrictive and a prior distribution that can be tuned in a continuous way between the two extremes is desired.
This can be achieved with the introduction of a ground state order parameter, $\eta(\vec{w})$, that is defined to be zero for ECI vectors, $\vec{w}$, that reside within the cone corresponding to the assumed ground states and positive outside of this cone. 
By then multiplying the Gaussian prior of Eq. \ref{eq:prior_distribution} by $e^{-\gamma\eta(\vec{w})}$ according to
\begin{equation}
    P(\vec{w}) \sim \prod^{K}_{k} \mathcal{N}(w_k|\mu_k,\alpha_k^{-1}) e^{-\gamma\eta(\vec{w})}
    \label{eq:ground_state_prior}
\end{equation}
it is possible to smoothly transition between the two extreme distributions by varying $\gamma$ between 0 and $\infty$. 
Indeed, when $\gamma=0$ the Gaussian prior of ridge regression, Eq. \ref{eq:prior_distribution}, is recovered since $e^{-\gamma\eta(\vec{w})}=1$.
In the other limit, $\gamma \rightarrow \infty$, the exponential, $e^{-\gamma\eta(\vec{w})}$ is equal to 1 inside the ground state cone (since $\eta(\vec{w})=0$ inside the cone), but goes to zero outside the cone (where $\eta(\vec{w}) > 0$).
In this limit, only ECI vectors $\vec{w}$ within the ground state cone of interest have a non-zero probability. 
To be a meaningful prior distribution for intermediate values of $\gamma$ (i.e. $0 < \gamma < \infty$), the ground state order parameter $\eta(\vec{w})$ should not only be positive for ECI vectors $\vec{w}$ outside the ground state cone of interest, but should decrease monotonically as $\vec{w}$ approaches the cone. 
This will ensure that the prior distribution of Eq. \ref{eq:ground_state_prior} for $0 < \gamma < \infty$ favors neighboring ground state cones more than ground state cones that are far way from the ``true" ground state cone in ECI space. 

\begin{figure}\label{fig:hull_envelope}
    \centering
    \includegraphics[width=1\linewidth,height=1\textheight,keepaspectratio]{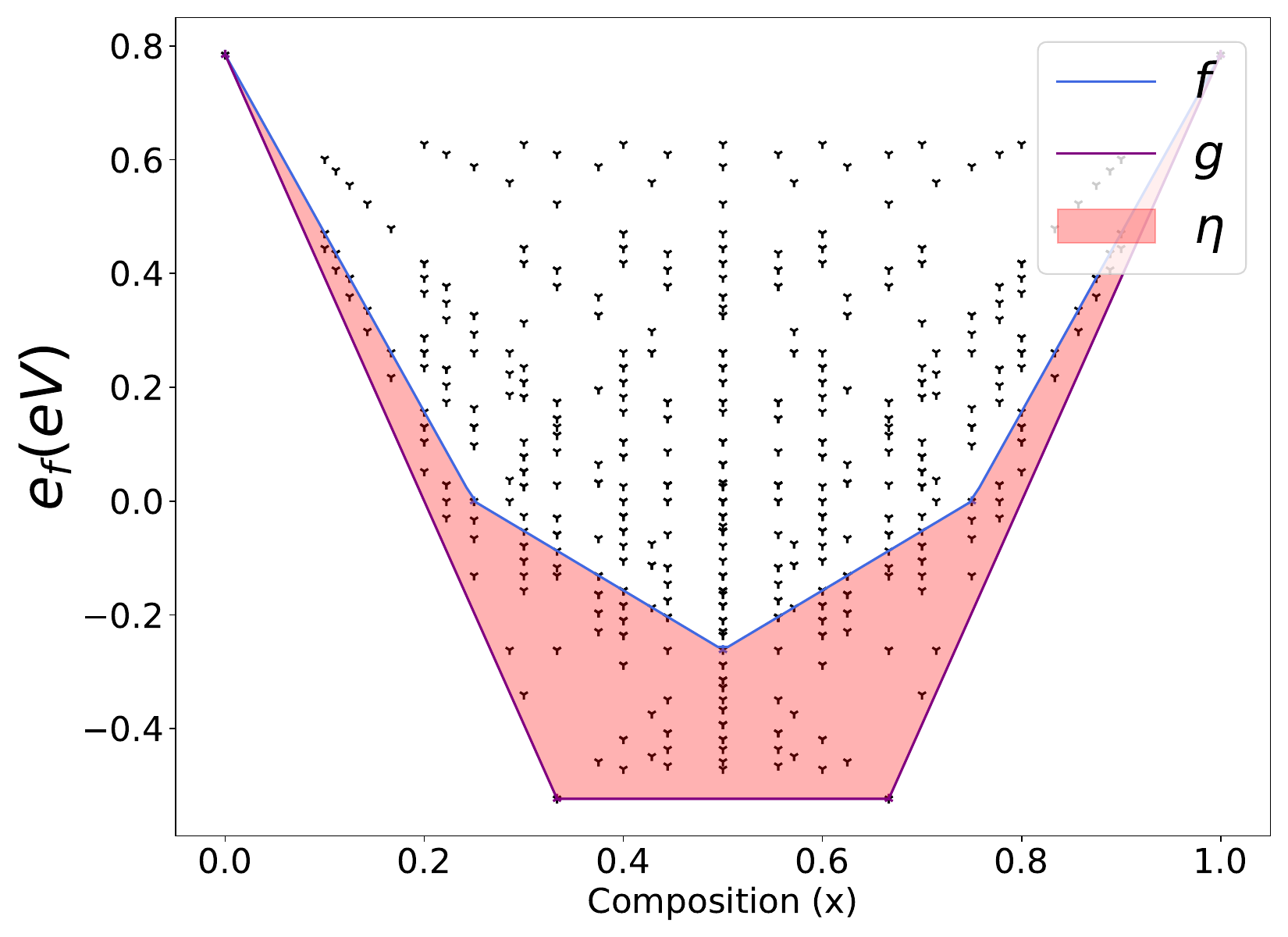}
    \caption{Depiction of the hull envelope function for a synthetic set of predicted formation energies, generated by an ECI vector in nearest neighbor, and second nearest neighbor space for the triangular lattice. ECI vectors within the ``true" cone will produce f and g curves that overlap, producing zero area between curves. }
    \label{fig:order_parameter_definition}
\end{figure}

A suitable ground state order parameter that satisfies the above constraints can be defined according to 
\begin{equation}\label{continuous_gsa}
     \eta(\vec{w}) =  \int_{0}^{1} f(\vec{w},x) - g(\vec{w},x)dx
\end{equation}
where $g(\vec{w},x)$ represents the piece-wise linear function that connects the ground states predicted by $\vec{w}$ as a function of composition $x$, while $f(\vec{w},x)$ represents the piece-wise linear function that connects the energies of the ``true" ground states. 
As schematically illustrated in Figure \ref{fig:order_parameter_definition}, $\eta(\vec{w})$ measures the area between the two piece-wise linear functions. 
The value of the integral in  \ref{continuous_gsa} is strictly positive because the predicted convex hull $g(\vec{w},x)$ is always less than or equal to  $f(\vec{w},x)$.
If $\vec{w}$ resides within the ground state cone of interest, then the predicted ground states are the same as the desired ground states and $\eta(\vec{w})=0$ since $g(\vec{w},x)=f(\vec{w},x)$. 
If $\vec{w}$ resides outside the ground state cone of interest, the value of $\eta(\vec{w})$ increases as $\vec{w}$ moves away from the cone of interest.

With a prior distribution of the form of Eq. \ref{eq:ground_state_prior}, the posterior distribution, assuming a Gaussian likelihood function (Eq. \ref{eq:likelihood_function}), can be expressed as
\begin{equation}
    P(\vec{w}|\vec{t},X) \sim e^{-\beta K(\vec{w},\lambda,\delta)}
    \label{eq:posterior_distribution}
\end{equation}
where 
\begin{equation}\label{eq:generalized_kost_function}
    K(\vec{w},\lambda) = |\vec{t}-X\vec{w}|^2 + \lambda |\vec{w}|^2 + \delta \eta(\vec{w}).
\end{equation} 
As before, $\beta$ is the precision appearing in the likelihood function (i.e. $\beta=1/\sigma_{L}^{2}$), $\lambda = \alpha/\beta$ is the regularization coefficient of ridge regression (assuming the same precision $\alpha$ for each non-truncated ECI) and $\delta=\gamma/\beta$ is a new parameter that measures the prior knowledge about the ground states.
When $\delta =0$ the above posterior distribution reduces to that of ridge regression, while $\delta \rightarrow \infty$ restricts posterior distribution to the ``true" ground state cone. 

The irregular shapes of ground state cones in ECI space will generally make it difficult if not impossible to construct an analytical expression for posterior distributions that rely on prior knowledge about the ground states. 
An alternative is to sample ECI vectors from a posterior distribution using Monte Carlo methods.
A Metropolis-Hastings Monte Carlo algorithm to sample ECI vectors from the posterior distribution of Eq. \ref{eq:posterior_distribution} can take the form:
\vspace{5mm}
\begin{algorithmic}
    \State $\Vec{w} \gets \vec{w}_{init}$
    \While{$n < N $}
        \State $\Vec{w}' \gets \vec{w} + \Delta\vec{w}$
        \State $r \gets e^{-\beta[\,K(\Vec{w}', \lambda, \delta)-K(\Vec{w}, \lambda,\delta)]\,}$             \If{$u \in [0,1]  < r$}
                \State $\Vec{w} \gets \Vec{w}'$
            \EndIf
        \State $n \gets n+1$
    \EndWhile
    \State $return \gets \vec{w}$
\end{algorithmic}
\vspace{5mm}
In this algorithm, a new ECI vector is generated by perturbing an existing ECI vector and accepting it with a probability proportional to the ratio of their posterior probabilities. 
This is repeated $N$ times, where $N$ is of the order of $10^4-10^5$. 
A subset of sampled ECI vectors (of the order of 100), taken at periodic intervals to reduce correlations, are then sent to downstream thermodynamic calculations. 
For example, each sampled ECI vector can be used to calculate an order-disorder transition temperature of one of the ground states using thermodynamic Monte Carlo simulations. 
Since the above Monte Carlo algorithm samples ECI vectors according to the posterior probability distribution of Eq. \ref{eq:posterior_distribution}, downstream thermodynamic predictions such as order-disorder transition temperatures can be averaged directly.

\subsubsection{Determining the probability of different ground state sets}\label{sec:cone_probability}

The ground states of a cluster expansion determine the topology of the temperature versus composition phase diagram. 
Any two ECI vectors sampled from the same ground state cone will generate phase diagrams that are topologically the same, differing only in the quantitative values of order-disorder transition temperatures and phase boundaries.
ECI vectors sampled from a posterior distribution that does not use prior knowledge about the ground states will not always predict the same ground state set and hence will generate phase diagrams belonging to different topology classes. 
In this context, it is of interest to quantify the probability of a particular ground state set and therefore of a particular phase diagram topology given DFT training data. 
This can be done by considering a ridge-regression posterior distribution and integrating the posterior probability within a particular ground state cone according to 
\begin{equation}
    P(C_h)=\frac{\int_{\vec{w}\in C_h}e^{-\beta \tilde{K}(\vec{w},\lambda)}d\vec{w}}{\int_{\vec{w}}e^{-\beta \tilde{K}(\vec{w},\lambda)}d\vec{w}}
    \label{eq:cone_probability}
\end{equation}
where the integral in the numerator is restricted to the volume of a ground state cone $C_h$ corresponding to a ground state set labeled $h$ and the integral in the denominator extends over the full ECI space.
In Eq. \ref{eq:cone_probability}, $\tilde{K}(\vec{w},\lambda)=|t-X\vec{w}|^2  + \lambda |\vec{w}|^2$ is the cost function of the ridge-regression posterior distribution. 
This probability assigns a weight to a particular topology class for a ridge-regression posterior distribution.

Equation \ref{eq:cone_probability} can be calculated with the introduction of a partition function and a corresponding free energy and by subsequently employing free energy integration techniques commonly used in statistical mechanics.
A convenient partition function over ECI space can be defined according to
\begin{equation}
    Z(\beta,\lambda,\delta)=\int_{\vec{w}}e^{-\beta K(\vec{w},\lambda,\delta)}d\vec{w}
\end{equation}
with a corresponding free energy $-\beta\Phi=\ln Z(\beta,\lambda,\delta)$. 
With this definition, the numerator in Eq. \ref{eq:cone_probability} becomes equal to $Z(\beta,\lambda,\delta=\infty)$, while the denominator in Eq. \ref{eq:cone_probability} becomes equal to $Z(\beta,\lambda,\delta=0)$.
The probability of Eq. \ref{eq:cone_probability} can then be expressed in terms of free energies according to 
\begin{equation}
    P(C_h)=\frac{Z(\beta,\lambda,\delta=\infty)}{Z(\beta,\lambda,\delta = 0)}=e^{-\beta[\Phi(\delta=\infty)-\Phi(\delta=0)]}
    \label{eq:cone_probability2}
\end{equation}
The free energies in turn can be calculated by exploiting the following derivative
\begin{equation}
    \frac{\partial \Phi}{\partial \delta}=-\frac{1}{\beta}\frac{1}{Z}\frac{\partial Z}{\partial \delta}
\end{equation}
which can be written more explicitly as
\begin{equation}
    \frac{\partial \Phi }{\partial \delta} = \frac{1}{Z} \int_{\vec{w}} \eta(\vec{w}) e^{-\beta K(\vec{w},\lambda,\delta)} d\vec{w} = \left<\eta(\delta)\right>
    \label{eq:free_energy_derivative}
\end{equation}
where $\left<\eta(\delta)\right>$ is the average of the ground state order parameter as a function of $\delta$ according to the posterior distribution of Eq. \ref{eq:posterior_distribution}.
This average can be calculated with the Monte Carlo algorithm introduced in the previous section over a grid of $\delta$ values. 
By integrating Eq. \ref{eq:free_energy_derivative} from $\delta=0$ to $\delta \rightarrow \infty$, it is possible to calculate the free energy difference appearing in Eq. \ref{eq:cone_probability2} according to
\begin{equation}
    \Phi(\delta=\infty)-\Phi(\delta=0) = \int_{0}^{\infty} \left<\eta(\delta)\right>d\delta
    \label{eq:free_energy_integral}
\end{equation}
By comparing the probabilities $P(C_h)$ for different ground state sets $h$, it is possible to assign relative probabilities for competing ground state sets, and therefore, for different phase diagram topologies.

\section{Results}
We compare two Bayesian approaches to parameterizing alloy cluster expansions: (i) the Bayesian interpretation of cross validated ridge regression described in Section \ref{sec:bayesian_fitting} and (ii) the Bayesian approach that uses knowledge about the ground states of the alloy as a prior introduced in Section \ref{sec:Bayesian_ground_state_priors}.
We use the rocksalt ZrN$_x$ phase, which exhibits substantial nitrogen-vacancy disorder, as a model system to test the two Bayesian approaches to constructing cluster expansions. 

\subsection{Training data and ground states}
The first step to constructing a cluster expansion for an alloy such as rocksalt ZrN$_x$ is to generate training data in the form of formation energies as calculated with a first-principles electronic structure method such as Density Functional Theory (DFT).
Different nitrogen-vacancy orderings on octahedral sites of the FCC Zr sublattice in the rocksalt form of ZrN$_x$ were generated using the CASM software package.\cite{puchala2023casm,puchala2023casmMonte} 
The energies of 1297 of these configurations were then calculated using DFT within the generalized gradient approximation of Perdew–Burke-Ernzerhof (PBE). \cite{perdew1996generalized} 
The DFT-PBE calculations were performed with the VASP package \cite{kresse1993ab,kresse1994ab,kresse1996efficiency,kresse1996efficient} using the projector augmented wave (PAW) pseudopotential method.\cite{blochl1994projector,kresse1999ultrasoft} The PAW potentials include the 4$s$, 4$p$, 5$s$ and 4$d$ orbitals of Zr (Zr sv) and the 2$s$, 3$p$ orbitals of N as valence states. 
A plane wave energy cutoff of 525 eV and an automatic k-point mesh with $R_k$ parameter of 45\AA (corresponding to a primitive cell k-point grid of 17x17x17) were used. 

The 1297 formation energies, calculated with Equation (\ref{formation_energy_equation}), are shown in Figure \ref{fig:convex_hull}.
These 1297 formation energies form the vector of training data $\vec{t}$ with which to parameterize cluster-expansion Hamiltonians, as described above. 
The rocksalt form of ZrN$_x$ has many nitrogen-vacancy ground state orderings, as is evident in the formation energy plot of Figure \ref{fig:convex_hull}.
The formation energy of each ground state is labeled with a Greek letter. 
Many of the ground states have complex nitrogen-vacancy orderings with periodicities defined by large supercells of the primitive FCC Zr unit cell. 
A large cluster expansion basis set is therefore necessary to be able to accurately reproduce the DFT ground states.
Some ground states are more stable than others. 
The more stable states will tend to disorder at higher temperatures and are therefore of more importance in determining properties of a material. 
The ground state at x=1/2, labeled as the $\beta$ phase, is especially stable, having a formation energy (Figure \ref{fig:convex_hull}) that is significantly lower than that of other competing ordered phases. 
In contrast, other orderings at higher concentrations (e.g. phases $\gamma$, $\delta$, $\varepsilon$) barely break the convex hull and are almost degenerate with many other nitrogen-vacancy orderings having similar nitrogen concentrations. 

\subsection{Bayesian Ridge Regression}
\label{sec:bayesian_ridge_regression}

\begin{figure}[h!]
    \centering
\includegraphics[width=1\linewidth,height=1\textheight,keepaspectratio]{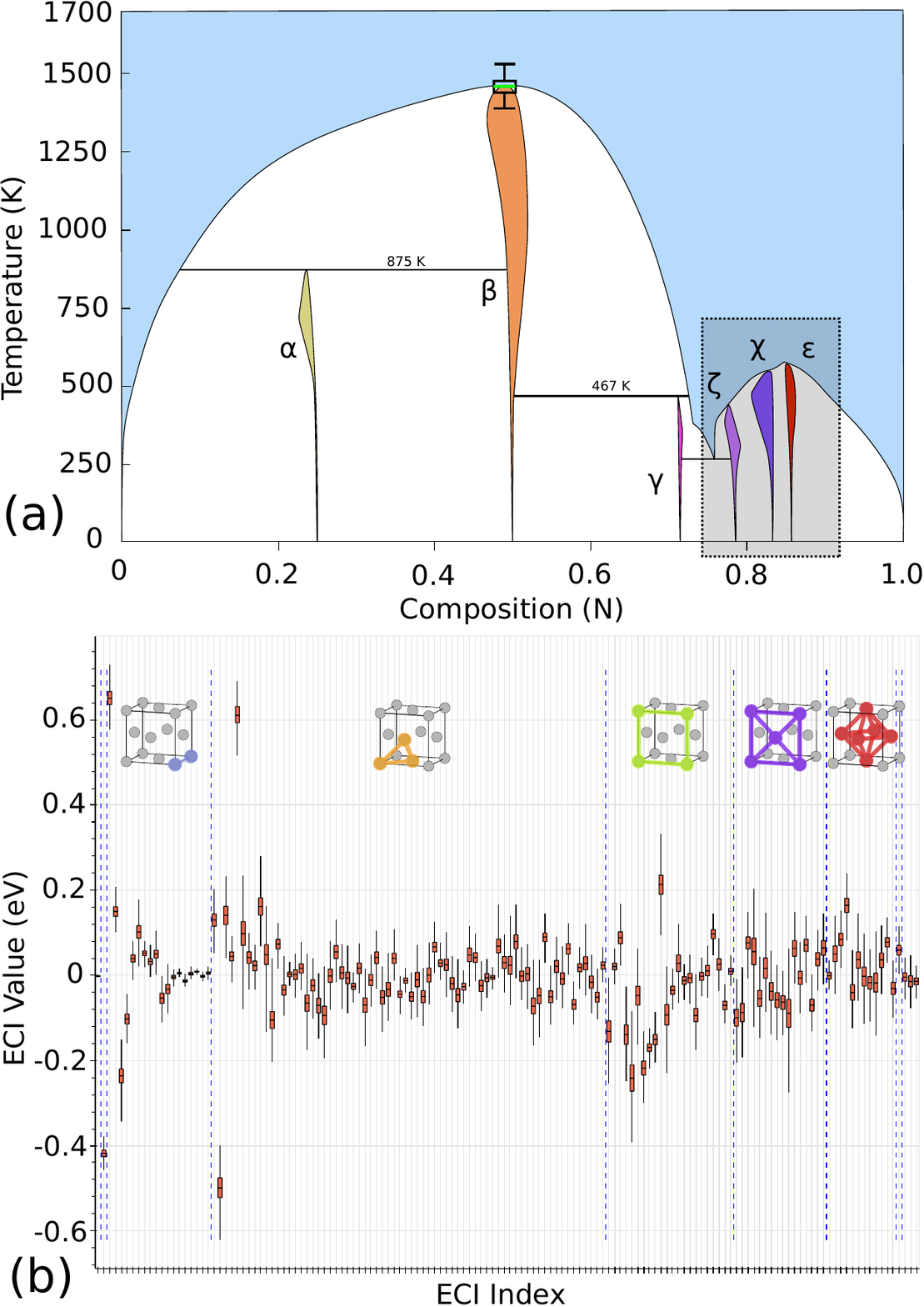}
\caption{(a):Order-disorder phase diagram of rocksalt ZrN$_x$, as calculated with the posterior mean cluster expansion using a mean zero (ridge-derived) prior. The large light-blue region is a disordered solid solution. Other colored regions, labeled by Greek letters, are stable compounds in which nitrogen and vacancies order over the octahedral interstitial sites of an FCC Zr sublattice. This phase diagram is missing the DFT-observed $\zeta$ ground state, and instead has the spurious ground state $\chi$. The gray region encompassing $\zeta$, $\chi$ and $\varepsilon$ represents the fact that the ridge-informed posterior mean model finds many possible new ground states in this composition region during Monte Carlo, and that the set of ground states in this region are not well defined (Hence, the motivation for the cone prior method). (b): Boxplots of ECI value vs ECI index for the Ridge-informed posterior distribution. Each ECI is depicted with its median value, as well as first and third quartiles as the lower and upper bounds of each box. The whiskers below and above boxes are the lowest and highest values for each ECI. ECI are grouped by cluster size from left to right. ECI of a given cluster size are bounded by vertical dashed lines. ECI values include the multiplicity of the ECI orbit.  }
    \label{fig:ridge_phase_diagram}
\end{figure}

In Bayesian ridge regression, the likelihood and prior functions, $P(\vec{t}|\vec{w},X,\beta)$ and $P(\vec{w}|\alpha)$, respectively, are modeled as Gaussians, where $\beta=1/\sigma_L^2$ is the precision on the training data and $\alpha=1/\sigma_{P}^2$ is the precision on the ECI in the prior. 
The maximum {\it a posteriori} (MAP) cluster expansion that maximizes the posterior Bayesian ridge distribution, $P(\vec{w}|\vec{t},X,\beta,\alpha) \sim P(\vec{t}|\vec{w},X,\beta)P(\vec{w}|\alpha)$, depends only on the ratio $\lambda =\alpha/\beta$, known as the ridge-regression regularization constant.\cite{Bishop2006} 
In cross-validated ridge regression, a value of $\lambda$ is chosen such that it minimizes a k-fold cross-validation score. 
This is a common approach with which cluster expansions are trained to DFT formation energies of different chemical orderings over the sites of a parent crystal structure.\cite{Mueller2009}

The 1297 formation energies of the rocksalt ZrN$_x$ phase shown in Figure \ref{fig:convex_hull} were used to generate a Bayesian ridge regression MAP cluster expansion containing 142 expansion coefficients. 
While the resulting cluster expansion has a small root mean square error (RMSE) between DFT and predicted formation energies of 3.7 meV/primitive cell, it predicts a qualitatively incorrect convex hull, with one missing ground state and one spurious ground state. 
The cluster expansion was subsequently implemented in Monte Carlo simulations to calculate a temperature versus nitrogen composition phase diagram using the CASM code. \cite{puchala2023casm} 
A Monte Carlo simulation cell size containing $24\times24\times24$ unit cells was used.
Thermodynamic integration techniques applied to grand canonical Monte Carlo data \cite{van2002self,puchala2023casmMonte} were used to calculate the free energies of the high temperature solid solution and the ground state phases. 

Figure \ref{fig:ridge_phase_diagram} shows the phase diagram as calculated with the Bayesian ridge MAP cluster expansion. 
The phase diagram contains six ground states that remain stable to elevated temperatures. 
The ground state labeled $\chi$ at a composition of  $0.8\bar3$ is a spurious ground state. 
Furthermore, since $\chi$ is part of the training data set whose energy was calculated with DFT, it is known to be a false ground state. 
The MAP cluster expansion also predicts additional ground states identified during slow cooling Monte Carlo simulations using a $24\times24\times24$ simulation cell. 
These low energy configurations are incommensurate with all supercells containing 16 or fewer primitive unit cells. 
Due to their complexity, we were unable to identify the supercell and basis ordering of the low energy configurations generated during cooling Monte Carlo simulations.
Most of the new ground states are stable between $x=0.75$ and $x=0.9$, highlighted in gray in Figure \ref{fig:ridge_phase_diagram}. 
This composition interval encompasses the $\zeta, \chi$ and $\varepsilon$ phases. 
The true equilibrium orderings and their order-disorder transition temperatures for the Bayesian ridge MAP cluster expansion are therefore unknown in that composition range.

Confidence intervals on order-disorder transition temperatures can be estimated once the posterior distribution, $P(\vec{w}|\vec{t},X,\beta,\alpha)$, of the cluster expansion is known. 
Each sampled ECI vector from the posterior distribution defines its own phase diagram, with distinct order-disorder transition temperatures. 
By sampling a large number of cluster expansions from the posterior distribution, and calculating their finite temperature thermodynamic properties using Monte Carlo simulations, it is possible to obtain a sample mean and standard deviation on transition temperatures. 
This, however, is only meaningful if all sampled cluster expansions predict the same ground states. 
While the ECI sampled from the Bayesian ridge posterior distribution have spurious  ground states, the ground state at $x=1/2$, labeled $\beta$, is also a DFT ground state and is very stable relative to other orderings at the same composition. 
To assess the uncertainty in the order-disorder temperature, we therefore focused on the $x=1/2$ ground state. 
We sampled 100 ECI vectors from the posterior ridge-regression distribution and subjected them to thermodynamic Monte Carlo simulations. 
A value for the precision on the DFT training data, $\beta=1/\sigma_L^2$, was estimated using a standard deviation of $\sigma_L\approx 4 meV/atom$, which is of the order of the k-point and energy cutoff convergence error of the DFT calculations. 
A value for the precision $\alpha=1/\sigma_P^2$ of the ECI prior distribution was estimated using the cross-validation optimized ridge-regression parameter according to $\alpha=\lambda\beta$. 
The uncertainty in the order-disorder transition is shown as an error bar near the top of the $x=\frac{1}{2}$ order disorder transition. 
The uncertainty on the order-disorder transition boundary is depicted in Figure  \ref{fig:ridge_phase_diagram} as a boxplot. The line within the central box is the median of all hundred order-disorder transition temperatures, with a value of 1457K. This value is slightly lower, but very similar to the order-disorder transition temperature of 1465K as predicted with the posterior mean cluster expansion.
The central box has lower and upper bounds of 1435K to 1473K respectively. These are also referred to as the first and third quartiles of the dataset, understood as the medians of the lower and upper halves of the transition temperatures. 
The lower and upper ``whiskers" extending below and above the box are lowest and highest values for transition. These have values of 1383K and 1528K respectively.

\subsection{Bayesian with ground state prior}
\label{sec:results_Bayesian_ground_state_prior}

The Bayesian ridge regression approach of the previous section, which uses a zero mean prior on the ECI, does not guarantee that sampled cluster expansions have a consistent set of ground states nor a set of ground states that are the same as those predicted with DFT. 
In cases where there is a high degree of certainty about the ground states, it is desirable to a embed that knowledge as part of the prior of a Bayesian posterior distribution. 
The Monte Carlo method introduced in Section \ref{sec:Bayesian_ground_state_priors} enables the sampling of ECI from a Bayesian posterior distribution with a prior distribution that is finite and bounded within the ground state cone of interest and zero outside of this cone. 
Every sampled ECI is then guaranteed to have the same ground states. 

We used the same set of 1297 DFT formation energies of the rocksalt ZrN$_x$ phase and the same value for the precision $\beta=1/\sigma_L^2$ on the DFT training data as used in the previous section for the Bayesian ridge regression cluster expansions.
The Monte Carlo method to sample ECI requires a way to check whether the sampled ECI $\vec{w}$ predicts the correct ground state set. 
To this end, we enumerated all symmetrically unique orderings of nitrogen and vacancies over the octahedral interstitial sites of rocksalt ZrN$_x$ up to supercell volumes containing 16 primitive unit cells. 
This results in approximately 247,000 unique configurations. 
Every sampled ECI $\vec{w}$ is then used to calculate the formation energies of the enumerated 247,000 configurations. 
If the ground states upon calculating the convex hull to the 247,000 formation energies differ from those of the ``true" ground states, the sampled ECI vector is rejected. 
The sampled ECI vectors that remain in the ground state cone of interest are accepted according to the algorithm described in Section \ref{sec:Bayesian_ground_state_priors}.
All accepted ECI vectors are guaranteed to predict the ``true" ground states and to occur with a frequency given by the Bayesian posterior distribution generated by multiplying a Gaussian likelihood with a prior that is finite and bounded within the cone of interest and zero outside that cone. 
The average of the accepted ECI vectors is therefore an approximation of the posterior mean cluster expansion. 
The posterior mean was again found to have a low RMSE of 3.8 meV per primitive cell, but in contrast to the Bayesian ridge method has no missing or spurious ground states. 

\begin{figure}[h!]
    \centering
\includegraphics[width=1\linewidth,height=1\textheight,keepaspectratio]{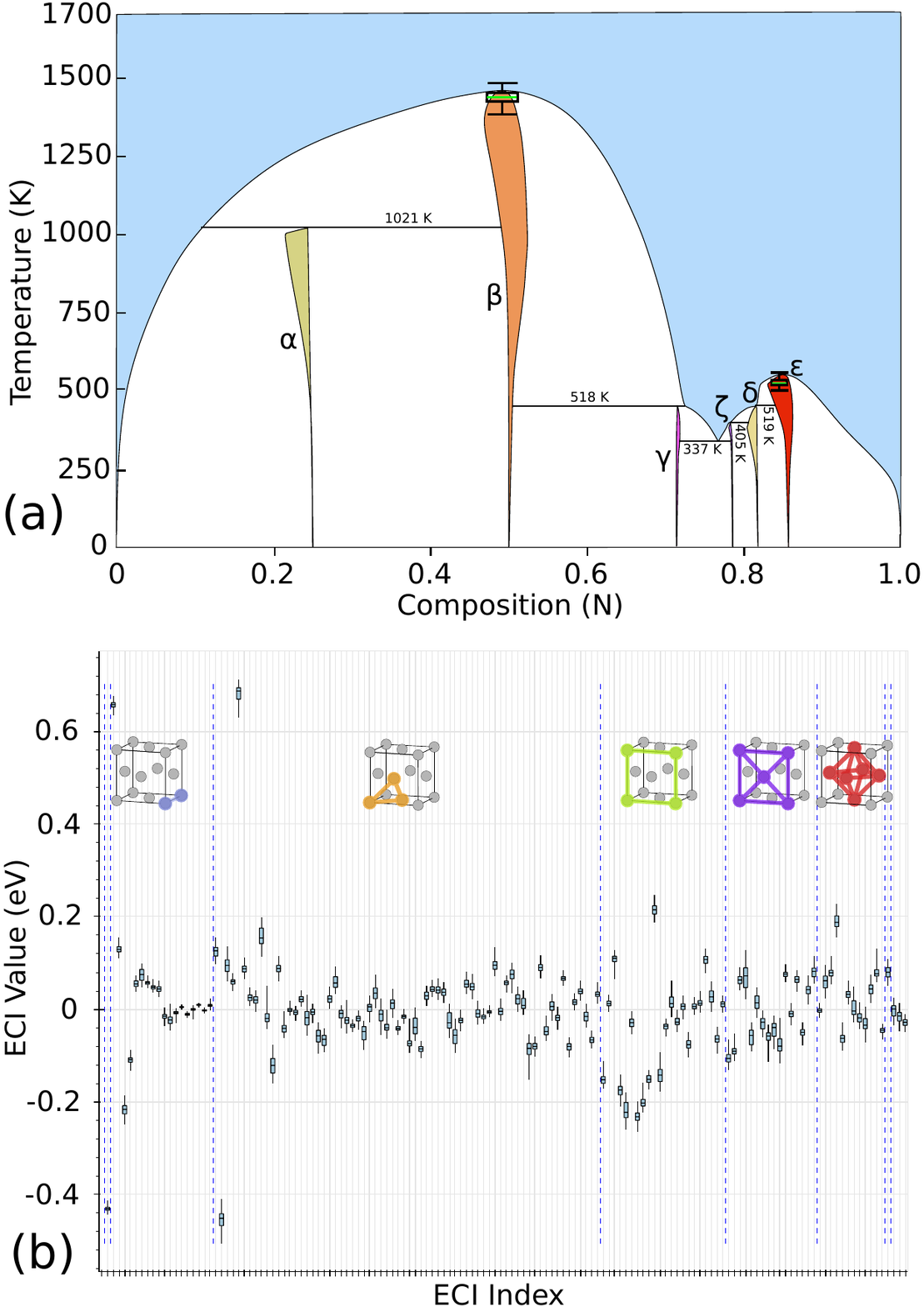}
\caption{(a): Order-disorder phase diagram of rocksalt ZrN$_x$, calculated with the cone-derived posterior mean cluster expansion. The large light-blue region is a disordered solid solution. Other colored regions, labeled by Greek letters, are stable compounds in which nitrogen and vacancies order over the octahedral interstitial sites of an FCC Zr sublattice. All DFT-predicted ground states ($\alpha,\beta,\gamma,\zeta,\delta, \varepsilon$) are replicated in this phase diagram, and there are no spurious ground states. Phases $\beta$ and $\varepsilon$ undergo congruent order-disorder transitions, and have uncertainty bounds on their transition temperatures that were estimated using 100 posterior cluster expansion samples. (b): Boxplots of ECI values vs ECI index for the cone-derived posterior distribution. Each ECI is depicted with its median value, as well as first and third quartiles as the lower and upper bounds of each box. Whiskers below and above boxes are the lowest and highest values for each ECI. ECI are grouped by increasing cluster size, from left to right. ECI of a given cluster size are bounded by vertical dashed lines. ECI values include the multiplicity of the ECI orbit.  }
    \label{fig:cone_post_phase_diagram}
\end{figure}

The calculated phase diagram using the posterior mean, ground state compliant cluster expansion is shown in Figure \ref{fig:cone_post_phase_diagram} and also contains six ground state orderings that remain stable at elevated temperatures.
Two of the ground states ($\beta$ and $\varepsilon$) undergo congruent order-disorder transitions to a high temperature solid solution, while the others ($\alpha$,  $\gamma$, $\zeta$ and $\delta$) decompose through peritectoid reactions upon heating.  
The ground state at x=1/2, labeled $\beta$, again has the highest order-disorder transition temperature, consistent with the fact that it is a very stable ordering. This is evident in the formation energy plot in Figure \ref{fig:convex_hull}. 
The other ground states at higher concentrations are weakly stable on the convex hull, are close in energy to many other orderings at the same concentration, and have much lower order-disorder transition temperatures. 
We point out that the phase diagram at low to intermediate nitrogen concentrations is a metastable phase diagram since pure Zr adopts the HCP crystal structure at low temperature and the BCC crystal structure at high temperature. 
Most importantly, this phase diagram correctly replicates all DFT-calculated ground states, without any spurious ground states, either from incorrect predictions of DFT-calculated configurations, or from new ground states found through Monte Carlo. 

To directly compare with the Gaussian ridge regression results, we again examined the uncertainty in the x=1/2 $\beta$ phase. 
This time, the posterior mean model predicts an order-disorder transition temperature for the $\beta$ phase at 1460 K. 
This is very close to the previous ridge-derived transition temperature of 1465 K. 
Again, the uncertainty in order-disorder transition temperature is depicted using a box plot. This time, the median transition temperature is 1447K, the first and third quartiles are 1434K and 1461K, and the lower and upper whiskers are at 1394K and 1494K. The uncertainty in transition temperature is reduced relative to the case of Bayesian Ridge regression, likely because the volume of accessible ECI values is much smaller when restricted to a single ground state cone. 

Now that the posterior distribution preserves ground states across all sampled models, it is possible to observe uncertainty intervals across the entire phase diagram. In a brute-force approach, one could now calculate an entire phase diagram for each sampled posterior ECI vector, and collect uncertainty bounds for each. This is possible, but requires significant computational resources. 
Ideally, we would only compute order-disorder transitions for all ordered phases. However, only $\beta$ and $\varepsilon$ exhibit congruent order-disorder transition, while all other phases disorder through peritectoids. It is computationally feasible to calculate temperatures for congruent transitions, which require repeated calculations of only one thermodynamic path. However, incongruent disorder through peritectoids involves a three-phase coexistence. Pinpointing this requires a fine mesh of thermodynamic paths for each posterior ECI sample, and would quickly regress to near-full phase diagram calculation for each sampled posterior model.
After sampling 100 phase diagrams from the posterior distribution, the $\varepsilon$ phase has a median  order-disorder transition temperature of 532K, with first and third quartiles of 526K and 542K, as well as lower and upper whiskers of 508K and 565K. Meanwhile, the posterior mean phase diagram predicts a transition temperature of 553K, seen as the boundary between the red $\varepsilon$ phase and the light-blue disordered phase in Figure \ref{fig:cone_post_phase_diagram}. 

\subsection{Probabilities of different ground state sets}

The ground state set predicted by a cluster expansion surrogate model determines the topology of the temperature versus composition phase diagram. 
This is evident upon comparing the phase diagrams of Figures \ref{fig:ridge_phase_diagram} and \ref{fig:cone_post_phase_diagram}. 
The two phase diagrams contain different phases as being stable at finite temperature since they were calculated with cluster expansions that predict different ground state sets. 
The example highlights a subtlety in the selection of a cluster expansion surrogate model for downstream thermodynamic calculations and the importance of taking the predicted ground state set of the model into consideration as part of model selection. 
As demonstrated in Section \ref{sec:results_Bayesian_ground_state_prior}, the new approach developed in Section \ref{sec:Bayesian_ground_state_priors} makes it possible to sample cluster expansions from a Bayesian posterior distribution that has a finite probability in a single ground state set cone and that is zero everywhere else in ECI space. 
With this new ability to restrict sampled cluster expansions to particular ground state set cones, a question that naturally arises is: What are the relative probabilities between competing ground state sets (and hence phase diagram topology classes) given the training data?

As outlined in Section \ref{sec:cone_probability}, free energy techniques from statistical mechanics can be applied to the Bayesian posterior distribution introduced in Section \ref{sec:Bayesian_ground_state_priors} to calculate the probabilities of each ground state set cone in ECI space. 
To develop a sense for the values of the probabilities of different ground state sets and, therefore, phase diagram topology classes, we calculated the probabilities for two ground state sets using a Bayesian posterior distribution in which the ridge regression parameter, $\lambda$, is set to zero. 
Setting $\lambda$ and $\delta$ equal to zero in Equations \ref{eq:posterior_distribution} and \ref{eq:generalized_kost_function} reduces the posterior distribution to that of the likelihood function (Eq. \ref{eq:likelihood_function}) and its maximum {\it a posteriori} cluster expansion is then simply the least squares fit, also referred to as the maximum likelihood model. 
Similar to the ridge-regression MAP cluster expansion of Section \ref{sec:bayesian_ridge_regression}, the least squares model for the ZrN$_x$ system resides in a ground state cone that differs from that predicted by the DFT data. 
We refer to the set of ground states predicted by the maximum likelihood (least squares) cluster expansion as $h_{ML}$ and its corresponding cone in ECI space as $C_{ML}$.
Similarly, we refer to the ground states predicted by the DFT data as $h_{DFT}$ and its corresponding cone in ECI space as $C_{DFT}$.
The probabilities $P(C_{ML})$ and $P(C_{DFT})$, as defined by Eq. \ref{eq:cone_probability}, represent the probabilities of the ground state sets $h_{ML}$ and $h_{DFT}$ given the training data and a variance of $1/\beta$ on each data point.
We calculated $P(C_{ML})$ and $P(C_{DFT})$ using Equations \ref{eq:cone_probability2} and \ref{eq:free_energy_integral} after calculating the average ground state order parameters $\left<\eta(\delta)\right>$ for $C_{ML}$ and $C_{DFT}$ as a function of $\delta$ using the Metropolis-Hastings algorithm introduced in Section \ref{sec:Bayesian_ground_state_priors}.
This proceedure predicts that $P(C_{ML})$ is approximately $4.1 \times 10^{-2}$ (roughly 4 \%), while $P(C_{DFT})$ is of the order of $1\times10^{-21}$. 
The low probability of $4.1 \times 10^{-2}$ for the maximum likelihood ground state set (which differs from the DFT ground state set) is surprising and suggests that many other ground state sets (and phase diagram topology classes) are consistent with the data. 
Even more surprising is the exceedingly low probability of the DFT ground state set.

The low probability of the DFT ground state set could have several sources. 
The probability $P(C_h)$ of residing within a particular ground state cone, $C_h$, according to the posterior probability distribution of Eq. \ref{eq:posterior_distribution} upon setting $\lambda$ and $\delta$ equal to zero depends on the volume and shape of the ground state cone and the likelihood function, Eq. \ref{eq:likelihood_function}, when evaluated within that volume. 
A small cone volume $C_h$ will translate into a small probability $P(C_h)$. 
$P(C_h)$ will also be small if the cone $C_h$ is far from the maximum likelihood model in ECI space, as the cone will then overlap with the distant tails of the likelihood function. 
The likelihood function, approximated as a product of iid Gaussian distributions, one for each data point, becomes sharper and narrower with increasing training data.\cite{Bishop2006}
Finally, it may also be possible that the numerical accuracy of the DFT calculations is insufficient to resolve the true ground states.

\section{Discussion}
Bayesian statistics provide a rigorous framework for quantifying and propagating uncertainties of surrogate models. 
In statistical mechanics, surrogate models are used to interpolate and generalize expensive first-principles electronic structure calculations within Monte Carlo simulations. 
The alloy cluster expansion  \cite{sanchez1984generalized,de1994cluster,van2018first} in particular is commonly used to calculate thermodynamic properties of solids that arise from the configurational degrees of freedom of distributing different chemical species over the sites of a parent crystal structure. 
While multiple studies \cite{Mueller2009,cockayne2010building,cluster_expansion_compressive_sensing,GOIRI2018257,KRISTENSEN2014,ALDEGUNDE2016} have cast the parameterization of a cluster expansion in a Bayesian context, only two \cite{KRISTENSEN2014,ALDEGUNDE2016} have leveraged the Bayesian framework for uncertainty quantification and propagation.  
In previous implementations of Bayesian cluster expansions,  prior precision has either been leveraged as a means of regularizing and coupling expansion coefficients \cite{Mueller2009}, or selected \cite{ALDEGUNDE2016} using the Relevance Vector Machine (RVM) \cite{tipping2001sparse}, which is based on a widely accepted statistical learning method known as the evidence approximation. \cite{Bishop2006} 

This contribution formulates a Bayesian approach for cluster expansion parameterization that uniquely incorporates the ground state problem into the prior selection process. 
The ground states of an alloy ultimately determine the topology of the temperature versus composition phase diagram, as each ground state in a three dimensional crystal will remain stable at finite temperature.
Each cluster expansion model, as represented by a vector of effective cluster interactions (ECIs), predicts a specific set of ground states. The union of cluster expansion models that predict the same set of ground states form cones in ECI space. 
This property makes it possible to explicitly enforce ground state replication within a Bayesian framework by restricting the prior probability distribution to the cone in ECI space corresponding to the ground states of interest. 
Often it is desirable to generate models that predict ground states consistent with the first-principles electronic structure method predictions. 

The ability to enforce a consistent set of ground states when sampling cluster expansion models from a Bayesian posterior distribution is essential to enable meaningful uncertainty quantification on calculated thermodynamic and kinetic properties. 
A common approach of establishing uncertainty bounds is to sample a large number of surrogate models from a Bayesian posterior distribution, using each to calculate a downstream thermodynamic property for the purpose of estimating averages and variances. 
Such an approach, when used to calculate uncertainties on the phase boundaries of a phase diagram, is only meaningful if the sampled surrogate models all predict a phase diagram belonging to the same topology class. 
The Bayesian formalism and accompanying Monte Carlo sampling approach introduced in this work provide a statistically meaningful way of restricting sampled cluster expansion models to all predict a consistent set of ground states. 
Furthermore, the approach makes it possible to assign weights to different phase diagram topology classes.
In cases when there is uncertainty about the correct ground states, Bayesian posterior distributions that are constrained to particular ground state sets enable hypothesis testing and allow an assessment as to which ground state sets have a higher density in the unconstrained posterior distribution compared to other ground state sets.

Finally, we comment on the types of errors that are being quantified and propagated in Bayesian approaches of cluster expansion parameterization. 
A first-principles statistical mechanics study of a particular alloy system will have several sources of errors and uncertainties.
The initial source of uncertainty is in the calculation of first-principles energies, which are used to train surrogate model Hamiltonians. 
These typically use a method based on DFT. 
Numerically tractable implementations of DFT rely on approximations, such as the local density approximation (LDA) or the generalized gradient approximation (GGA), and results can vary greatly depending on the approximation.\cite{cohen2012challenges,verma2020status} 
The Bayesian approaches described in this paper do not quantify this type of uncertainty, instead assuming that the particular approximation to DFT used to generate training data is the ground truth for the surrogate model. 

Within a choice of a particular first-principles electronic structure method, uncertainty also stems from numerical convergence tolerances. 
Any form of DFT software requires some truncated basis to describe electron wavefunctions. 
For methods using a plane wave basis, the plane wave energy cutoff determines basis truncation and is one source of numeric uncertainty.\cite{lejaeghere2016reproducibility} 
Likewise, periodic simulation cells can be solved in reciprocal space, requiring a grid of wave vectors $\vec{k}$, with the k-point grid density being another source of numerical uncertainty. 
These, and any other user-selected thresholds are uncertainty sources which can be quantified and propagated. 
Within the Bayesian frameworks described here, this uncertainty is estimated with the likelihood precision $\beta$.

Once first-principles energies are obtained, the data is approximated with a cluster expansion model (section  \ref{cluster_expansion_section}). 
Typically, a small number of expansion coefficients replicate most of the observed DFT data, and the remaining expansion coefficients can be set to zero (i.e. truncated).
Aside from specific scenarios entailing long-range interactions (Coulombic interactions and coherency strain \cite{laks1992efficient,wolverton2000first,wolverton2000short,blum2004mixed,van2007first,wang2023generalization,behara2024chemomechanics}), the most significant atomic interactions involve short-range, few-body interactions. 
For these reasons, truncating the number of cluster expansion terms is usually justifiable. 
Truncation from an infinite to a finite set of expansion terms is another source of uncertainty. 
Within a Bayesian framework, truncation is addressed in the prior distribution, where each truncated ECI is assigned a delta-function prior distribution function.

\section{Conclusion}
This work uses discontinuities in observed thermodynamic properties as constraints on surrogate models for use in first-principles thermodynamic calculations. 
These constraints are formalized as Bayesian priors, allowing for uncertainty quantification when calculating finite temperature thermodynamic properties in multi-component solids. 
Furthermore, this work also shows that encoding thermodynamic constraints through a Bayesian prior is an ideal approach for high throughput uncertainty quantification, as it ensures that propagated thermodynamic quantities share the same topological structure and are therefore comparable. 
While the cluster expansion is easily cast in a Bayesian context, the idea of Bayesian surrogate models with thermodynamically informed priors is a general paradigm; the concept of identifying qualitative discontinuities in observed data and using these discontinuities as constraints on surrogate models is general and should apply broadly to statistical mechanics modeling.

\section{Acknowledgments}
We would like to acknowledge helpful discussions with Prof. Anirudh Natarajan, Prof. Yu-Xiang Wang and Prof. M. Scott Shell.
This work was supported by the U.S. Department of Energy, Office of Basic Energy Sciences, Division of Materials Sciences and Engineering under Award \#DE-SC0008637 as part of the Center for
Predictive Integrated Structural Materials Science (PRISMS Center).
We are grateful for computing resources from the National Energy Research Scientific Computing Center (NERSC), a U.S. Department of Energy Office of Science User Facility, operated under Contract No. DE-AC02-05CH11231 using NERSC award BES-ERCAP0023147. Use was made of computational facilities purchased with funds from the National Science Foundation (CNS-1725797) and administered by the Center for Scientific Computing (CSC). The CSC is supported by the California NanoSystems Institute and the Materials Research Science and Engineering Center (MRSEC; NSF DMR 2308708) at UC Santa Barbara. 

\clearpage

\bibliography{citations}
\end{document}